\documentclass[a4paper,twocolumn,fontsize=10pt,DIV=16,abstract=true]{scrartcl}
\pdfoutput=1

\usepackage[utf8]{inputenc}
\usepackage[T1]{fontenc}
\usepackage[lighttt]{lmodern}
\usepackage[protrusion=true,expansion=true]{microtype}

\usepackage[style=ACM-Reference-Format, backend=biber]{biblatex}
\usepackage{authblk}
\usepackage{graphicx}
\usepackage{xcolor}
\usepackage{hyperref}
\usepackage[nameinlink]{cleveref}
\usepackage{caption}
\usepackage{subcaption}
\usepackage{float}
\usepackage{booktabs}
\usepackage{multirow}
\usepackage{bytefield}
\usepackage{listings}

\crefname{lstlisting}{listing}{listings}
\Crefname{lstlisting}{Listing}{Listings}

\hypersetup{
  colorlinks,
  linkcolor={black},
  citecolor={green!50!black},
  urlcolor={blue!50!black}
}

\begin{document}

\title{Advancing Protocol Diversity\\in Network Security Monitoring}
\subtitle{A Refined Software Architecture and Implementation\\for Efficient Modular Packet-Level Analysis}

\author[1]{Jan Grashöfer}
\author[2]{Peter Oettig}
\author[3]{Robin Sommer}
\author[3]{Tim Wojtulewicz}
\author[1]{Hannes Hartenstein}

\affil[1]{Karlsruhe Institute of Technology,
KASTEL -- Institute of Information Security and Dependability\authorcr
\texttt{\{jan.grashoefer,hannes.hartenstein\}@kit.edu}}

\affil[2]{Karlsruhe Institute of Technology,
Steinbuch Centre for Computing (SCC)\authorcr
\texttt{peter.oettig@kit.edu}}

\affil[3]{Corelight, Inc.\authorcr
\texttt{\{robin,tim\}@corelight.com}}

\date{}

\maketitle

\begin{abstract}
With information technology entering new fields and levels of deployment,
e.g., in areas of energy, mobility, and production, network security monitoring needs to be able to cope with those environments and their evolution.
However, state-of-the-art Network Security Monitors (NSMs)
typically lack the necessary flexibility to handle the diversity of the packet-oriented layers below the abstraction of TCP/IP connections.
In this work, we advance the software architecture of a network security monitor to facilitate the flexible integration of lower-layer protocol dissectors while maintaining required performance levels.
We proceed in three steps:
First, we identify the challenges for modular packet-level analysis, present a refined NSM architecture to address them and specify requirements for its implementation.
Second, we evaluate the performance of data structures to be used for protocol dispatching,
implement the proposed design into the popular open-source NSM Zeek and assess its impact on the monitor performance.
Our experiments show that hash-based data structures for dispatching introduce a significant overhead while array-based approaches qualify for practical application.
Finally, we demonstrate the benefits of the proposed architecture and implementation by migrating Zeek's previously hard-coded stack of link and internet layer protocols to the new interface.
Furthermore, we implement dissectors for non-IP based industrial communication protocols and leverage them to realize attack detection strategies from recent applied research.
We integrate the proposed architecture into the Zeek open-source project and publish the implementation to support the scientific community as well as practitioners,
promoting the transfer of research into practice.
\end{abstract}

\section{Introduction}
In network security monitoring, stateful deep packet inspection is used to passively obtain detailed information about the communication in a network.
This information allows operators to unveil direct and indirect security threats,
such as intrusion attempts and misconfigurations.
Network Security Monitors (NSMs) log summaries of their observations for forensic purposes,
generate alerts or trigger instant countermeasures upon detecting malicious behavior~\cite{hunt_network_2012}.

As information technology spreads into ever new domains,
trends like the Industrial Internet of Things (IIoT) lead to a diversification of the traditional TCP/IP-based protocol stack \cite{casola_security_2019, zhu_you_2020}.
For example, numerous works that address security in production \cite{kwiecien_profinet_2009, zihao_feng_snort_2016, wong_enhancing_2017, pfrang_detecting_2018} or energy systems \cite{kabir-querrec_cyber_2017, bohara_ed4gap_2020} show the large demand for deep packet inspection (DPI) of non-IP protocol stacks.
This need is also unabated with regard to the proliferation of machine learning,
as it has been shown that the knowledge of protocol semantics has a higher influence on the quality of trained models than the applied algorithm~\cite{anderson_machine_2017}.
However, existing research work that extends established open source monitoring software like Snort, Suricata or Zeek/Bro to support new low-level protocols,
is hardly ever transferred into practice%
\footnote{For example, the work of \citeauthor*{kabir-querrec_cyber_2017} \cite{kabir-querrec_cyber_2017}: \url{https://github.com/zeek/zeek/pull/76}}.
The integration of new dissection capabilities is severely hindered by the far-reaching changes these extensions require to the monitoring tools,
because the low-layer processing, which operates at the granularity of packets, is typically hard-coded, due to performance concerns.
Consequently, established monitoring solutions are limited to a narrow, TCP/IP-based protocol stack.

To advance protocol diversity in network security monitoring,
we develop a refined NSM architecture for efficient modular packet-level analysis.
Modular protocol dissectors have three main advantages:
1) Separating the core monitor functionality, i.e. the orchestration of the analysis,
from the protocol dissection significantly simplifies the maintenance of the monitor core as well as the implementation of new dissectors for external developers.
2) Providing dissectors as independent modules using well-defined interfaces facilitates their reuse in new contexts.
3) Modularity allows NSM operators to tailor the monitoring system to their deployment-specific needs.
As a monitor is confronted with arbitrary input that might even
be crafted to attack the NSM itself \cite{ptacek_insertion_1998, paxson_bro:_1999, sassaman_halting_2011},
the increased flexibility aids in minimizing the potential attack surface of the monitor.

Given the inherent complexity of dissector development and the ever growing number of protocols,
established tools already introduced plugin interfaces to decouple the domain-specific development of dissectors for application-layer protocols.
Protocol dissectors for protocols that build upon TCP/IP are able to exploit the existing interfaces.
Examples like DNP3~\cite{lin_adapting_2013} or Modbus dissectors are widely adopted, reused, and are even being continuously developed~\cite{hill_using_2019, chromik_parser_2019, udd_exploiting_2016}.
However, the existing plugin interfaces cannot be easily extended to support packet-level dissectors, as these interfaces focus on a different level of abstraction.
In this work, we introduce a refined software architecture for NSMs that addresses the unique challenges of packet-level analysis.
Among these, performance represents the key challenge for a dynamic plugin scheme compared to the previous, hard-coded processing at packet-level.
Continuous monitoring for applications like intrusion detection requires online-processing to guarantee timely reactions, in spite of high traffic volumes or resource-constrained monitoring devices.
For modular packet-level analysis, dispatching performance becomes crucial,
because dispatching of packet-level dissectors is done for \emph{every} layer of \emph{every} packet.

So far, the lack of protocol support beyond the traditional Ethernet-based TCP/IP stack prevents the application of established network security monitoring technologies to protect sensitive resources like critical infrastructures.
By developing a refined network security monitor architecture for efficient modular packet-level analysis, we seek to close this gap.
Our contributions can be summarized as follows:
\begin{itemize}
\item We identify the challenges for modular packet-level analysis in DPI,
present an extended NSM architecture to address them, and derive requirements for its implementation.
\item We evaluate the performance of data structures for dissector dispatching and show that array-based approaches outperform hash-based data structures in this case,
while arrays can keep up with hard-coded packet-level processing.
\item We implement the proposed architecture into the established Zeek network monitor, modularize its previously hard-coded packet-layer stack, and measure a negligible performance impact on the overall monitor performance.
Furthermore, we simplify dissector development by supporting the use of a tightly integrated parser generator,
which allows for adding new dissectors without writing any C++ code.
\item We demonstrate the utility of the extended architecture on the example of dissectors for industrial communication protocols, namely GOOSE and ProfinetIO,
by leveraging them to realize basic, domain-specific attack detection strategies as proposed in recent research.
\end{itemize}
The overall goal of this work is to allow for the application of established network security monitoring technologies in new domains and,
in particular, to support the transfer of current research into practice.
In addition to providing the discussed examples as open source,
our implementation of the extended NSM architecture has been released as part of Zeek 4.0.

The rest of the paper is structured as follows:
First, we discuss related work in \Cref{sec:related_work}.
Then, we provide background on NSM architecture in \Cref{sec:background}.
In \Cref{sec:design_and_implementation}, we propose an extended NSM architecture for modular packet-level analysis,
specify requirements for its implementation, and present our implementation in the open-source NSM Zeek.
In \Cref{sec:evaluation}, we evaluate the performance of data structures for dissector dispatching and assess the performance impact of the extended architecture on the overall monitor performance.
Finally, we showcase a set of practical applications to demonstrate the utility of a refined architecture for modular packet-level analysis in \Cref{sec:practical_applications}.
This covers the migration of Zeek's packet-level protocol stack,
the integration with the Spicy parser generator toolchain \cite{sommer_spicy:_2016},
the implementation and exemplary application of dissectors for ICS protocols,
as well as an additional feature to keep track of unknown protocols.
\Cref{sec:conclusion} concludes the paper.

\section{Related Work}
\label{sec:related_work}

The established open source security monitoring tools,
Snort, Suricata, and Zeek/Bro, support a modular interface for dissectors based on TCP or UDP.
However, only Snort offers interfaces to customize processing of lower layers.
Snort 2 supports so-called \emph{preprocessors} for the reassembly of streams and additional detection capabilities \cite{koziol_intrusion_2003}.
While this allows to process other protocols,
the integration with other components like the rule engine is very limited.
Preprocessors also ignore layering so that every packet is processed by every preprocessor.
Snort 3 refines preprocessing and introduces a maximum of $256$ modular \emph{codecs} that integrate into the processing pipeline focusing decapsulation and allow "basic per-frame validation" \cite{snort_project_snort3_2021}.
Internally, Snort 3 maintains its own mapping of identifiers to codecs that is partially hard-coded.
This approach requires the codecs to maintain an additional mapping between the protocol-specific identifiers and the ones used by Snort.
Furthermore, the selection of Snort-specific identifiers needs coordination between extension developers to prevent the duplicate use of an identifier.

The need for modularization of network monitoring was also recognized by \citeauthor*{casola_security_2019} \cite{casola_security_2019}.
They extend a commercial monitoring software to support modular protocol dissectors focusing on IoT and wireless sensor networks.
They implement the monitoring of the IPv6 over Low power Wireless Personal Area network (6LoWPAN) protocol.
While \citeauthor*{casola_security_2019} implement a modular system,
they neither provide details about their interface design,
nor evaluate the performance impact introduced by modularization.
In fact, the presented architecture uses distributed probes that capture the traffic,
encapsulate each packet, and forward the traffic to the actual monitor.
This suggests that the dissectors for network layer protocols are implemented on top of the existing IP analysis stack.
The approach to encapsulate low-layer protocols in IP-based protocols to work around the limitations of the existing plugin mechanisms is also found in other work.
For example, the Idaho National Laboratory (INL) released a protocol dissector
that requires a converter to support an industrial automation protocol,
which is primarily specified for serial communication \cite{cisa_icsnpp-bsap-serial_2021}.

The analysis of protocol headers is also a performance critical aspect for networking hardware,
because the headers encode information that network components like switches, routers, and firewalls rely on to make their forwarding decisions.
In their work \citetitle{gibb_design_2013},
\citeauthor*{gibb_design_2013} review the process of dissecting multilayered packets in hardware to obtain relevant data like addresses \cite{gibb_design_2013}.
The authors discuss general challenges of packet parsing, describe the notion of parse graphs as state machines, and define an abstract model of packet parsers.
Furthermore, \citeauthor*{gibb_design_2013} consider reconfigurability.
However, their work addresses a different domain by focusing on high-speed packet processing in hardware using ASICs.
With the advent of software defined networking,
flexibility in terms of parsing low-level protocols has become a major concern.
\citeauthor*{bosshart_p4_2014} introduced a domain-specific language for Programming Protocol-independent Packet Processors (P4) \cite{bosshart_p4_2014} that gained significant momentum.
P4 fosters hardware independent reconfiguration of network devices in the field to support new header formats.
While these works underline the need to support protocol diversity,
they do not extend to network security monitoring,
which focuses on the comprehensive reconstruction of the observed communication up to the application layer.

\section{Background on Network Security Monitor Architecture}
\label{sec:background}

A network security monitor faces two fundamental challenges to provide visibility into network communication:
First, descriptive information must be extracted from the network traffic using deep packet inspection.
Second, the extracted information must be processed with respect to operator-defined network policies (e.g., to raise alarms)
or persisted for retrospective analysis (e.g., in the form of logs).
In this work, we focus on the first aspect of extracting information from the observed traffic.
In general, the processing of traffic in a network security monitor can be divided into three levels of abstraction (c.f. \cite{li_vnids_2018}): Packet-level, session-level, and artifact-level.
At the \textbf{packet-level}, the monitor extracts information that is defined at the scope of a single packet such as Ethernet addresses.
Packets group information that is sent from A to B and meta data that is required to do so.
In this work, we focus on the packet-level of DPI.
At the \textbf{session-level}, extracted information is combined to identify sessions (e.g., 5-tuple for a TCP/IP connection) that maintain a state.
Sessions can be nested; for example, multiple HTTP sessions might reuse a persistent TCP/IP connection.
At the \textbf{artifact-level}, artifacts like files are reconstructed from potentially multiple sessions.
These levels of abstraction are reflected in the software architecture of a network security monitor.

\begin{figure}[t!]
  \centering
  \includegraphics[width=\columnwidth]{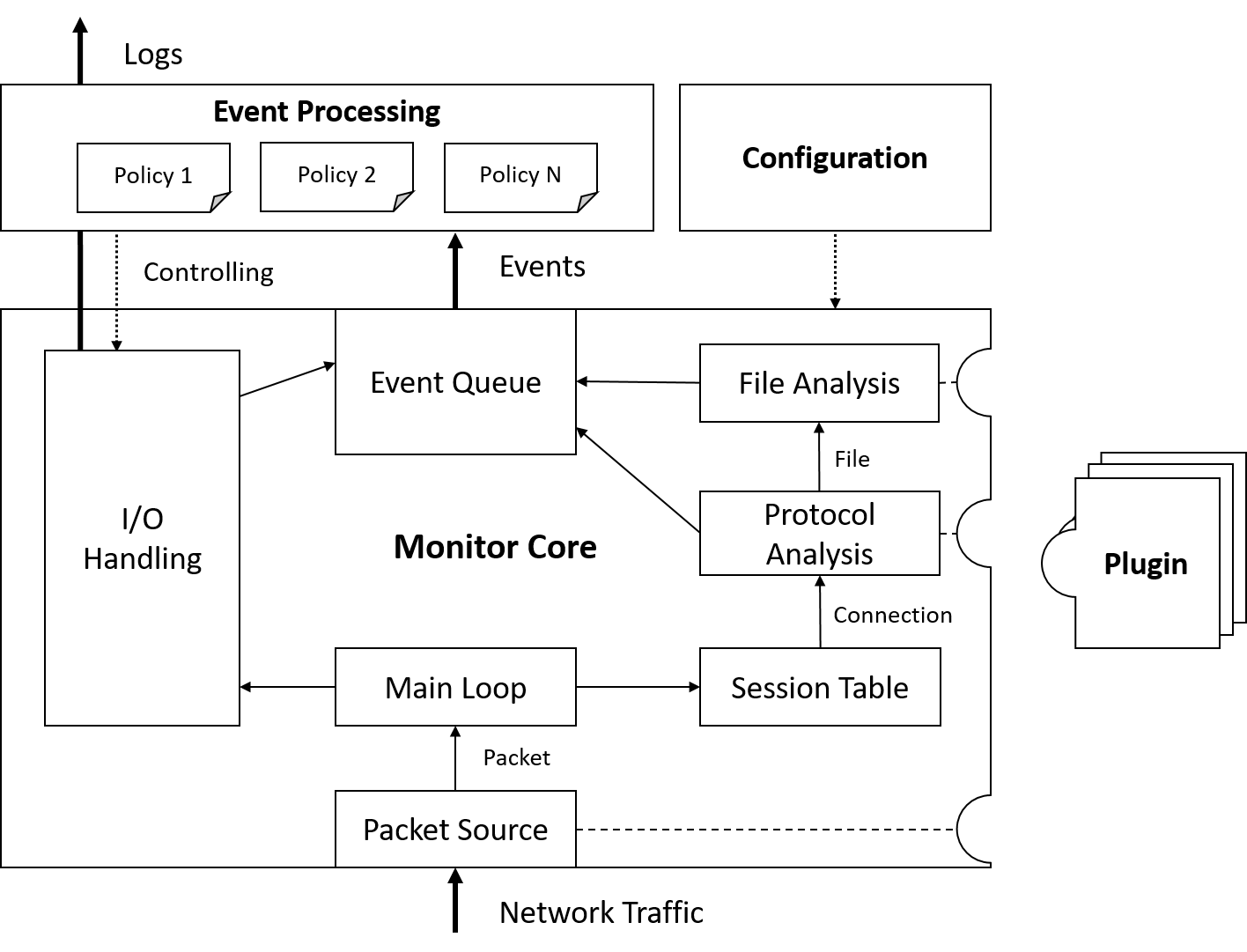}
  \caption{Network Monitor Reference Architecture\protect\footnotemark -- The monitor core transforms network traffic into a stream of high-level events that are processed based on user-defined policies, e.g., to control logging. The analysis performed by the monitor core can be customized and enriched by configuration and independently deployable plugins, respectively.}
  \label{fig:architecture}
\end{figure}

For our work, we define a network security monitor reference architecture as shown in \Cref{fig:architecture}.
We model our reference architecture based on Zeek/Bro \cite{paxson_bro:_1999},
as it proves to be sufficiently general to represent a variety of systems
\cite{stavrou_count_2014, de_carli_beyond_2014, li_vnids_2018}.
The key idea is to separate the DPI mechanism from the specification of policies by transforming the ingested network traffic into a stream of high-level events that describe the observed communication.
The monitor core is responsible for performing deep packet inspection to generate the event stream.
Users interact with the monitor core by specifying policies, providing configuration, and deploying plugins.
The user-defined policies act on the stream of high-level events.
Examples of events may be the establishment of a TCP connection or the observation of an HTTP header.
Depending on the capabilities of the employed policy language,
policies may be used for basic inspection and filtering (e.g., to control logging) up to the realization of attack detection logic by correlating multiple events.
The configuration as well as plugins can be used to customize the DPI process itself.
Using well-defined interfaces (represented as notches in \Cref{fig:architecture}),
plugins allow for the implementation of additional functionality separately from the monitor core,
which simplifies both,
the development process of plugins and the maintenance of the monitor core.
As plugins can be developed, compiled, and deployed without interfering with the NSM's code base,
NSM developer, plugin developer, and NSM user become independent roles.
We refer to the users of an NSM as operators.

\footnotetext{based on \citetitle{sommer_following_2017} \cite{sommer_following_2017}}%

In the reference architecture, the core's main loop is the central coordinator of the control flow in the monitor and steers the processing of packets and events.
Assume the monitor receives a packet that is part of an HTTP connection over TCP/IP.
The main loop reads the packet from a packet source.
Packet sources abstract different forms of packet ingestion,
such as reading from network interfaces or trace files, and are implemented as plugins.
A packet source is solely responsible for feeding the raw traffic into the NSM and does not involve parsing of the packet contents.
Once our example packet is read from the packet source,
the lower layers are processed to assemble the 5-tuple (source address and port, destination address and port, transport protocol) that is used to look up the corresponding TCP connection in the session table.
The session table maintains the state of active connections.
When the packet is assigned to a connection,
the analysis continues on the application-protocol-level, HTTP in our example.
If the protocol carries files, further file analysis may be conducted,
such as calculating file hashes.
Both protocol analysis and file analysis can emit events that are scheduled in the event queue.
Finally, the scheduled events are processed with respect to the user-defined policies.
Policies may control the handling of in- and output such as blocklists and logs as well as implement correlation and detection logic.

While the current architecture allows for easy extension of the monitor core by additional application-level protocol dissectors and file analyzers using the available plugin interfaces,
the processing of packet-level protocols is typically hard-coded.
In the following, we analyze the process of packet-level analysis and propose an extended architecture for modular packet-level analysis.

\section{Design and Implementation of Modular Packet-Level Analysis}
\label{sec:design_and_implementation}

In this section, we propose an extended NSM architecture for packet-level analysis,
describe the unique challenges at the packet-level, and derive the corresponding requirements for implementation.
Finally, we present our implementation of a framework for modular packet-level analysis in Zeek.
In \cref{sec:evaluation}, we evaluate the related performance aspects.

\subsection{Proposed Extended Architecture and Problem Analysis}
\label{sec:extened_architecture}

\begin{figure}[t!]
  \centering
  \includegraphics[width=\columnwidth]{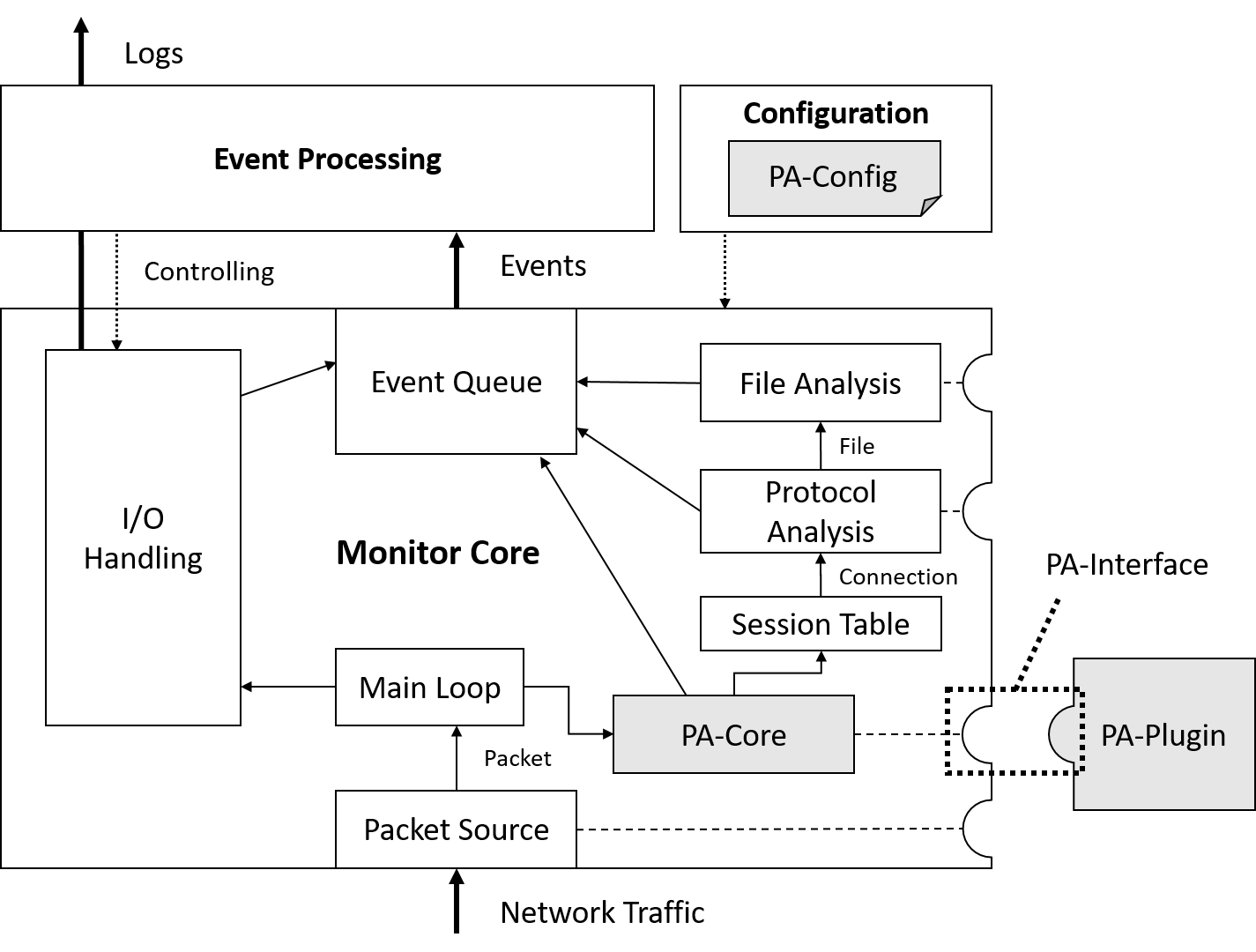}
  \caption{Proposed extended Network Security Monitor Architecture -- Packet-level analysis becomes a separate step and provides an interface for plugins. The configuration allows for orchestration of the plugins.}
  \label{fig:architecture_new}
\end{figure}

With respect to the reference NSM architecture as described in \Cref{sec:background},
modular packet-level analysis means to split the packet analysis code by protocol,
decouple protocol-specific dissectors from the monitor core by employing the plugin mechanism that facilitates extensibility, and enable operators to orchestrate the interaction between the dissectors.
\Cref{fig:architecture_new} shows our proposal for an extended network monitor architecture.
We introduce packet analysis as a separate processing step (PA-Core) in the monitor core.
The core part of our framework serves as the entry point for packet-level analysis
and defines the interface (PA-Interface) that is implemented by packet analysis plugins (PA-Plugin).
The PA-Interface defines how the PA-Core interacts with PA-Plugins to perform packet processing,
while the PA-Plugins realize the protocol dissection.
To allow for orchestration of the analysis process by operators,
we introduce the packet analysis configuration (PA-Config).

\begin{figure}[t!]
  \centering
  \begin{subfigure}[b]{0.6\columnwidth}
    \centering
    \includegraphics[width=\textwidth]{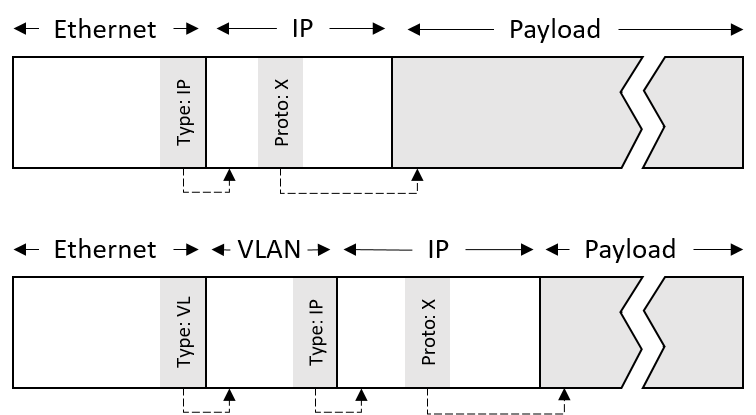}
    \caption{Example Packet Structures}
    \label{fig:packets}
  \end{subfigure}
  \hfill
  \begin{subfigure}[b]{0.35\columnwidth}
    \centering
    \includegraphics[width=\textwidth]{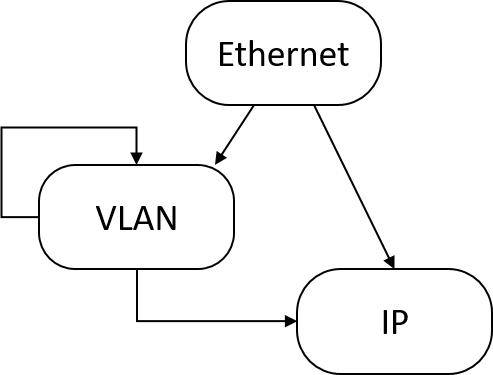}
    \caption{Protocol Transition Graph}
    \label{fig:ptg}
  \end{subfigure}
  \caption{Illustrative example of nested packet structures and a corresponding protocol transition graph inspired by \cite{gibb_design_2013}.}
  \label{fig:concepts}
\end{figure}

The actual challenge for the implementation of the extended architecture lies in the diversity and the nested structure of protocols in today's and future's networks
on the one hand and on the required performance of packet-level analysis on the other hand.
Let us recall that packet-level protocols are combined by encapsulating their Protocol Data Units (PDUs) into each other.
A PDU consists of its payload, i.e., data to be transferred,
and a header that contains meta data required to do so.
Encapsulating protocol Y into X means that Y's PDU becomes the payload of X's PDU,
so that the resulting packets consist of several layers represented by their protocol headers.
\Cref{fig:packets} shows two exemplary packet structures that carry IP over Ethernet with one of the packets using an 802.1Q VLAN tag.
In real-world environments, packets may consist of eight layers and more \cite{gibb_design_2013}.

On the packet-level, protocol headers typically contain a numerical value that identifies the encapsulated protocol.
The process of determining the encapsulated protocol and passing on the payload (i.e., the encapsulated PDU) for appropriate analysis, we call \textbf{packet-level dispatching}.
The mapping of protocol identifier to encapsulated protocol is part of a protocol's specification and, thus, might vary between protocols,
which makes dispatching context dependent.
Note that protocol identification on session-level is more complex \cite{dreger_dynamic_2006, grashofer_attacks_2020}.
However, protocol identification on session-level is usually done once per session,
whereas packet-level dispatching is required for \emph{every} layer of \emph{every} packet.
Thus, providing fast and context-aware dispatching represents the core challenge for implementing modular packet-level analysis.
In our extended architecture, the PA-Interface codifies how dispatching among PA-Plugins is realized.

In addition to the protocol formats, their interrelations need to be considered for dispatching,
i.e. which protocols can be encapsulated in a PDU.
Adapted from parse graphs defined by \citeauthor*{gibb_design_2013} \cite{gibb_design_2013},
we define Protocol Transition Graphs (PTGs) as digraphs for which nodes represent protocols and edges depict the possible encapsulation relationships.
\Cref{fig:ptg} shows the PTG for the packets shown in \ref{fig:packets}.
The graph contains a loop for the VLAN protocol to take into account stacked VLAN tags (QinQ, IEEE 802.1ad).
Parsing a packet layer by layer can be understood as executing the state machine that is described by the PTG.
The current state corresponds to the protocol of the currently analyzed layer.
State transitions are executed based on the identifier for the encapsulated protocol that is extracted from the current header.
Given the vast amount of protocols and the lack of a comprehensive registry%
\footnote{For example, the IANA lists more than 200 assignments for EtherTypes \cite{eastlake_ieee_2021}.
The list is not comprehensive because even a number of widely deployed protocols such as ProfinetIO, EtherCAT ore GOOSE are not officially registered.}
it is neither feasible nor desirable to support monitoring all of them.
For example, any dissector that is not required in a particular deployment context
unnecessarily increases the attack surface of the monitor.
Thus, operators also need the flexibility to adapt the packet-level analysis to their environment.
Customizing the analysis means to define the PTG state machine by selecting the relevant protocols and specifying the corresponding transitions.
We ensure this flexibility by splitting the analysis task into protocol-specific PA-Plugins and
moving the specification of the transitions into a separated PA-Config that can be adapted by the operator.

To summarize the problem addressed in this paper,
given the extended network monitor architecture of \Cref{fig:architecture_new} and the interrelations of the protocols to be analyzed on packet-level as sketched in \ref{fig:ptg},
one has to define the architectural elements, particularly the PA-Interface,
such that the performance penalty of the introduced additional functionality is minimal.
This problem, together with more detailed requirements, is addressed in the following.

\subsection{Requirements and Implementation}
\label{sec:requirements_and_implementation}

Following a structured approach, we decompose the problem of realizing modular packet-level analysis by deducing detailed requirements.
Subsequently, we present our implementation of the previously proposed extended architecture in Zeek,
and show that our implementation fulfills the defined functional requirements.

\paragraph{Requirements.} We define the following fundamental requirements for implementing flexible packet-level analysis:

\textbf{R1 -- Extensibility} of the NSM represents the key requirement and is enabled by modularity.
While modularity realizes decoupling of functionalities, extensibility allows to exploit the modularity at deployment time:
The addition of new protocol dissectors to an NSM deployment should be possible without modification of the monitor core.
Once dissectors can be developed and compiled independently from the NSM core,
the process of implementing new dissectors is vastly simplified while at the same time,
the maintainability of the monitor core is improved.

\textbf{R2 -- Universality} requires that the PA-Interface accounts for variability with respect to the location of protocol identifiers in the protocol header.
For example, assuming to find a protocol identifier at a fixed offset is not viable,
because the header structure might even vary within a protocol.
One prominent example is IPv6,
which uses the last of possibly multiple extension headers to identify the encapsulated protocol \cite{RFC8200}.

\textbf{R3 -- Independency} requires PA-Plugins to be able to schedule events in the event queue,
so that policies can be specified referring to packet-level information.
This is particularly important for protocols that do not establish sessions but carry relevant information like the Address Resolution Protocol (ARP) \cite{RFC0826} or protocols for industrial control systems.

\textbf{R4 -- Configurability} allows the operator of the NSM to select the set of analyzers and configure protocol identifier mappings without touching code.
On the one hand, this allows the operator to tailor the monitor to its environment and thus reduce the attack surface that the monitor itself offers.
On the other hand, configurability accounts for incomplete information.
For example, MPLS may not explicitly specify which protocol follows after a label \cite{RFC3032} so that the encapsulated protocol has to be inferred based on the deployment context.
With respect to the design of the PA-Config, there are two further requirements to consider:

\textbf{R4.1 -- Unambiguousness} refers to the fact that each protocol is free to define its own mapping between identifiers and encapsulated protocols.
Thus, the same identifier might refer to different encapsulated protocols depending on the context.
Conversely, there might be different identifiers for a single protocol.
For example, the Ethertype mapping used in the IEEE 802 standard family \cite{RFC7042} and the Point-to-Point (PPP) protocol field mapping \cite{RFC1661} exhibit both,
overlapping and multiple identifiers for the same protocol.

\textbf{R4.2 -- Integrability} ensures that the PA-Config allows for chaining dissectors arbitrarily to handle encapsulation.
In terms of a PTG this means that it is possible to establish arbitrary loops%
\footnote{Note that we assume each analysis step to consume a packet layer.
Thus, loops in the PTG cannot be exploited to block the progress of the analysis.}.
For example, data link layer PDUs might be encountered as payloads on higher layers with respect to tunneling protocols.
Furthermore, integrability ensures that existing dissectors or chains of dissectors can be reused.

\textbf{R5 --} In addition to the previously described functional requirements,
\textbf{Performance} is paramount.
In case of network security monitoring,
losing packets means to irretrievably lose potentially critical observations.
Thus performance is ultimately correlated to the visibility a monitor aims to provide.
Because nested layers trigger multiple packet-level dissectors for every packet that is seen on the wire,
the performance of the packet-level dispatching mechanism becomes a relevant factor.
We evaluate the dispatching performance in \cref{sec:evaluation}.
Note that apart from monitoring high volume traffic,
performance is also crucial with respect to the use of resource-constrained monitoring devices.
In the following, we present our implementation of the extended NSM architecture in Zeek and show that it satisfies the functional requirements.

\begin{figure}[t]
\centering
\includegraphics[width=0.9\columnwidth]{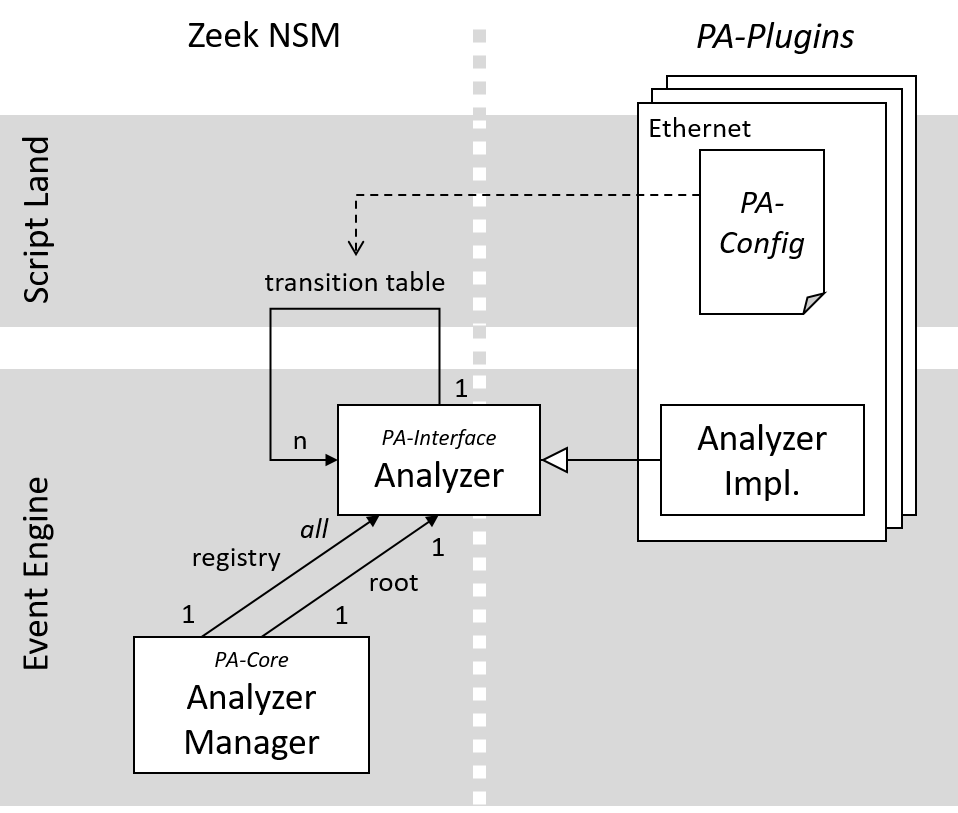}
\caption{Implementation of the extended NSM architecture for modular packet-level analysis in Zeek.}
\label{fig:design}
\end{figure}

\begin{table*}[t!]
    \centering

    \begin{tabular}{l l | p{7cm} l }
    \multicolumn{2}{c|}{\bfseries Data Structure} & \bfseries Description & \bfseries Implementation \\
    \toprule
    \multirow{2}{*}{Array}
     & Static & Array indexed by the protocol identifier & C++ built-in \\
     & Dynamic & Array of IDs trimmed at front and end & \lstinline$std::vector$ \\
    \hline
    \multirow{2}{*}{Tree}
     & Tree Map & Red/black tree & \lstinline$std::map$ \\
     & Array Tree & Tree grouping adjacent identifiers in arrays & custom \\
    \hline
    \multirow{4}{4em}{Hash Map}
     & Separate Chaining & Handle collisions using a list of items per index & \lstinline$std::unordered_map$ \\
     & Cuckoo Hashing & Calculate multiple candidate indices for an element using independent hash functions & open source \cite{brechko_cuckoo_2013} \\
     & Universal Hashing & Find a collision free hash function for a given input set from a family of hash functions & based on \cite{goos_efficient_1999} \\
     & Perfect Hashing & Construct a collision free hash function & based on \cite{fox_practical_1992, hanov_throw_2011} \\
    \bottomrule
    \end{tabular}

    \caption{Data structures evaluated for packet dispatching.}
    \label{tbl:data_structures}
\end{table*}

\paragraph{Implementation.}
The key idea for the implementation of the extended NSM architecture is to emulate the state machine that is described by a Protocol Transition Graph,
where parsing a packet layer by layer can be understood as executing the PTG state machine as described in \Cref{sec:background}.
\Cref{fig:design} shows an overview of our implementation in Zeek.
The monitor core is realized by Zeek's so-called event engine,
which is responsible for generating the high-level event stream.
Building upon the event engine,
Zeek comes with a turing-complete, domain-specific scripting language that acts as the primary user interface for NSM operators.
Scripts are used to express monitoring policies as well as to provide configuration for the event engine.
To allow for extension of the event engine,
Zeek offers a plugin mechanism based on shared libraries.

We define the PA-Interface in form of an abstract dissector super class, called \emph{analyzer}.
Derived packet analyzer implementations for a given protocol are provided by PA-Plugins.
Each analyzer implementation corresponds to a state of the PTG state machine.
The state machine's transition function is split across the different analyzers and represented in form of a transition table per analyzer.
Together with an analyzer implementation,
each PA-Plugin contains a corresponding PA-Config that specifies the transition tables using a policy script (see \ref{sec:traditional_stack} for an example).
The analyzer implementation specifies the actual dissection logic.
Typical steps include the verification of the PDU header and the extraction of the identifier that determines the encapsulated protocol.
Further processing might include scheduling events.
Finally, the current PDU's payload is forwarded based on the extracted protocol identifier using the transition table.
If no suitable analyzer is found, the failed dispatching is logged.
Otherwise, the analysis continues with the next analyzer.
Both evaluating the PA-Config and the dispatching mechanism are implemented as part of the analyzer super class, i.e. the PA-Interface,
as this functionality is shared between all analyzers.
The PA-Core is represented by the analyzer manager,
which maintains a registry of all available packet-level analyzers as well as a dedicated root analyzer.
In contrast to application-level analyzers that are spawned per connection,
each packet-level analyzer implementation is only instantiated once by the manager,
since on-demand creation would severely degrade performance.
The dedicated root analyzer serves as the entry point for the packet-level analysis in the event engine.

The described implementation achieves extensibility (R1) by splitting the analysis task per protocol and moving it into separately deployable PA-Plugins.
By placing the responsibility for identifier extraction on the analyzer,
the implementation also meets the universality (R2) requirement,
as we do not introduce restricting assumptions on the identifier placement.
Due to the modular design of the employed script interpreter,
PA-Plugins are able to define and emit custom events without the need to modify the NSM core.
Hence, the design satisfies the independency (R3) requirement as well.
Configurability (R4) is realized by placing the transition table definition into Zeek scripts that can be adapted by NSM operators to customize the packet-level analysis process.
The unambiguousness~(R4.1) and integrability (R4.2) requirements for the configuration are both addressed by the state-machine-driven approach for dispatching.
As each packet analyzer (i.e., state) maintains its own transition table,
identifier conflicts between different protocols are prevented.
However, this comes at the cost of storing multiple, potentially large transition tables.
The integrability requirement (R4.2) is satisfied,
since there are no restrictions on possible transitions.
For example, a new IP-carrying stack of data link protocols can be added,
without the need to reimplement any IP-related functionality.
Due to the paramount importance of performance (R5), we examine this aspect separately in \Cref{sec:evaluation},
including an evaluation of the design's memory overhead.
All in all, the presented implementation establishes a clear separation between the roles of NSM operators, NSM core developers, and the developers of packet-level analyzers.
The implementation is publicly available being merged into the BSD-licensed open source Zeek project%
\footnote{\url{https://github.com/zeek/zeek}}
and is part of its 4.0 release.

\section{Performance Evaluation}
\label{sec:evaluation}

This section is dedicated to the performance of the modular packet-level analysis approach as presented in the previous section.
During packet processing, dispatching requires to look up the analyzer for a given identifier and to forward payloads accordingly.
One of the main concerns is whether it is feasible to replace the hard-coded dispatching with a dynamic plugin scheme on such a critical data path.
Note that, for example, just a single download at the speed of 10 MB/s, with each packet consisting of 5 packet-level layers on average,
yields more than ten thousand lookups per second.
Thus, the dispatching data structure becomes mission-critical.
Our goal is to assess and minimize the potentially introduced overhead.
In this section, we first evaluate different data structures in the context of the dispatching use case.
In a second step, we investigate the impact of the proposed architecture on the overall performance of the monitor,
employing the best-performing data structure for our implementation in Zeek.
For reproducibility, we release our experiments and code as open-source%
\footnote{\url{https://github.com/kit-dsn/packet-analyzer-benchmarks}}.

\subsection{Dispatching Data Structures}
\label{sec:benchmark}

For dispatching, we require a simple data structure that maps protocol identifiers to packet analyzer instances.
\Cref{tbl:data_structures} provides an overview on the evaluated data structures.
In addition, we generate code that uses if or switch statements for comparison with hard-coded approaches.
To benchmark the possible data structures,
we use a dedicated benchmark application written in C++ that simulates the dispatching process using a single instance per data structure.
For our measurements we employ the Google Benchmark library%
\footnote{\url{https://github.com/google/benchmark}}.
We measure dispatching time, startup time, and consider memory usage as well as caching behavior for each data structure.
The startup time is the time required to build a data structure.
The dispatching time is the overall time spent for identifier lookups.
Startup time and memory usage solely depend on the configured identifier mappings.
The dispatching time is also influenced by the monitored network traffic,
as its structure determines the sequence of dispatching steps.
We describe the traffic structure using a simplified, CSV-based trace format:
Each line corresponds to a packet and each field to a layer,
with the values representing protocol identifiers.
All measurements are performed on a system equipped with a 2.6GHz Intel\textsuperscript{\textregistered} Core\textsuperscript{\textregistered} i7-6600U processor (2 cores, 64 KiB L1d cache, 512 KiB L2 cache) and 20 GB DDR4 memory.
The benchmark application was compiled using clang version 10.0.1.

In our benchmarks we consider both,
the traffic composition and the identifier mapping.
For each of these dimensions we simulate two scenarios,
one that aims to capture the characteristics of a real-world deployment and one that simulates a deliberately adverse situation.
Aiming to capture the properties of real-world traffic,
we obtain a trace from the Monday PCAP of the CIC-IDS17 data set \cite{sharafaldin_toward_2018}.
Following the CIC-IDS17 traffic,
the adverse trace consists of packets that contain three packet-level PDUs each,
but whose identifiers are randomly drawn from a uniform distribution in the range of $1$ to $10.000$.
In practice, a similar situation could occur due to incorrect parsing or being triggered intentionally by an attacker. 
For comparability both traces comprise a total number of 10 million PDUs to dispatch.
The real-world-oriented identifier mapping covers the protocols that would be supported in a default configuration of Zeek.
To account for a challenging case,
we use a deliberately fragmented mapping that covers hundred identifiers found in the randomized trace as well as the identifiers of the Zeek mapping.

\begin{figure}[t!]
  \centering
  \begin{subfigure}[b]{\columnwidth}
    \centering
    \includegraphics[width=0.99\textwidth]{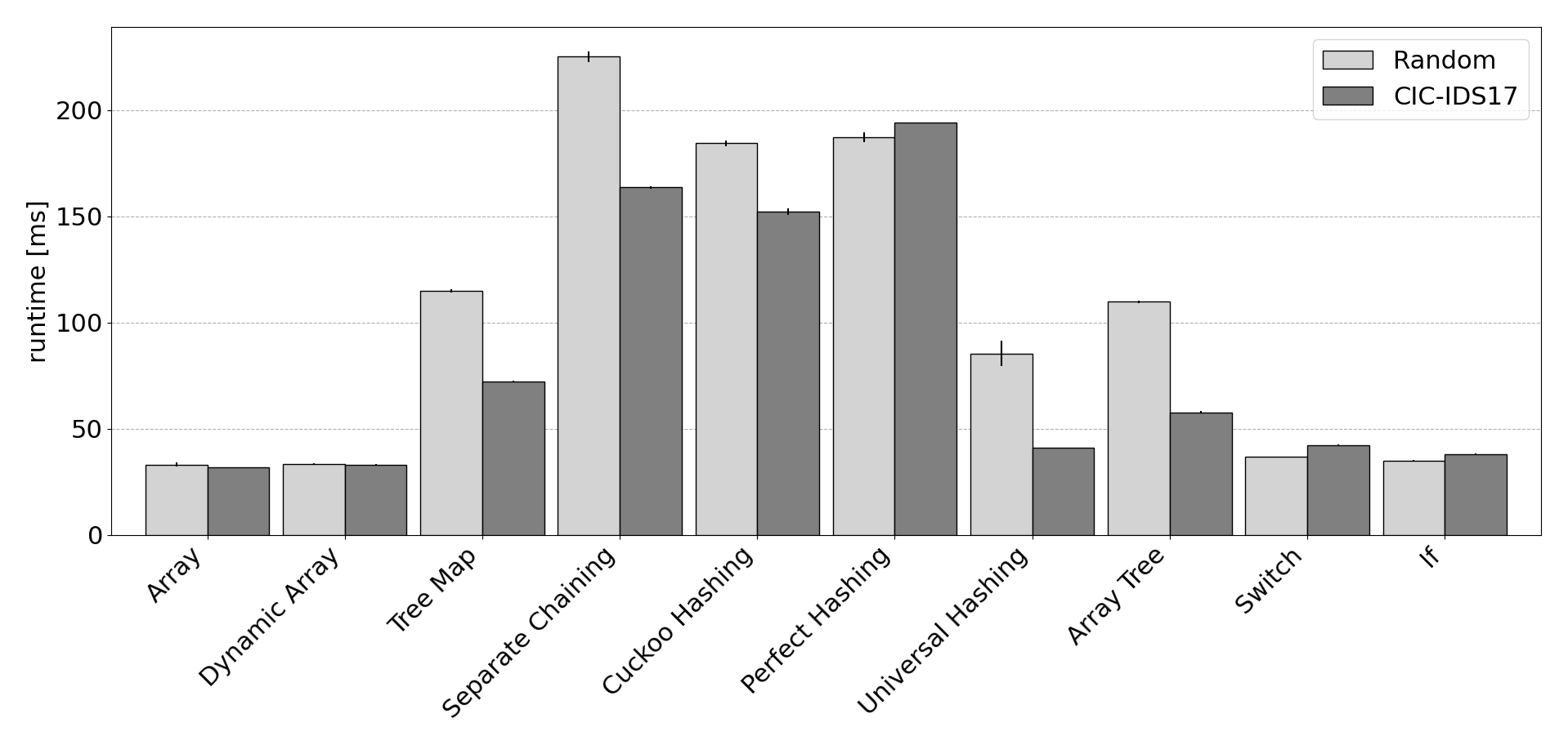}
    \caption{Concise identifier mapping based on Zeek defaults under CIC-IDS17 and randomized traffic composition.}
    \label{fig:eval_zeek}
  \end{subfigure}
  \begin{subfigure}[b]{\columnwidth}
    \centering
    \includegraphics[width=0.99\textwidth]{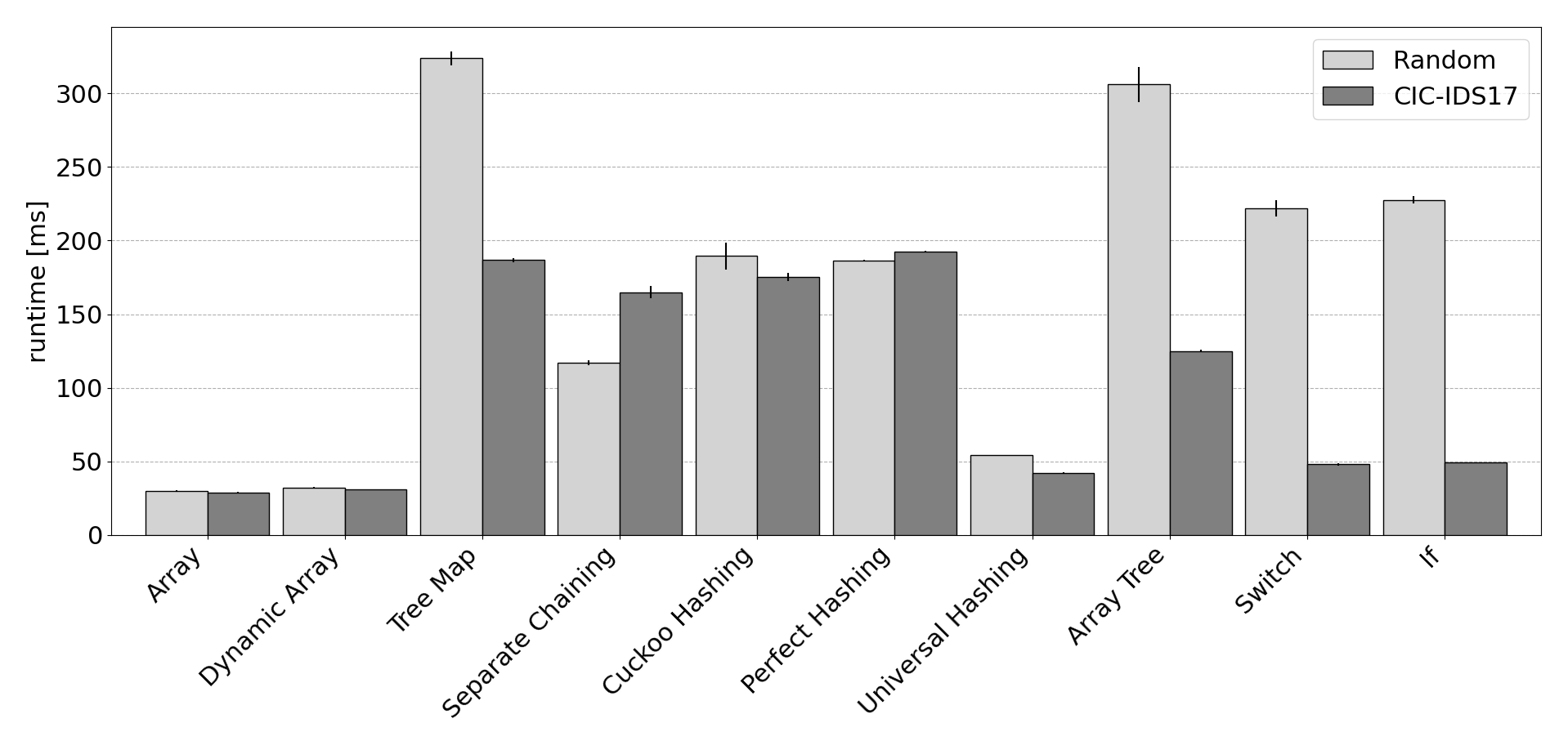}
    \caption{Fragmented identifier mapping (Zeek IDs plus 100 IDs from random trace) under CIC-IDS17 and randomized traffic composition.}
    \label{fig:eval_fragmented}
  \end{subfigure}
  \caption{Dispatching times for selected data structures. Each bar corresponds to the average of 10 runs. Error bars depict the 95\% confidence intervals.}
  \label{fig:eval_times}
\end{figure}

\begin{figure}[t!]
  \centering
  \includegraphics[width=\columnwidth]{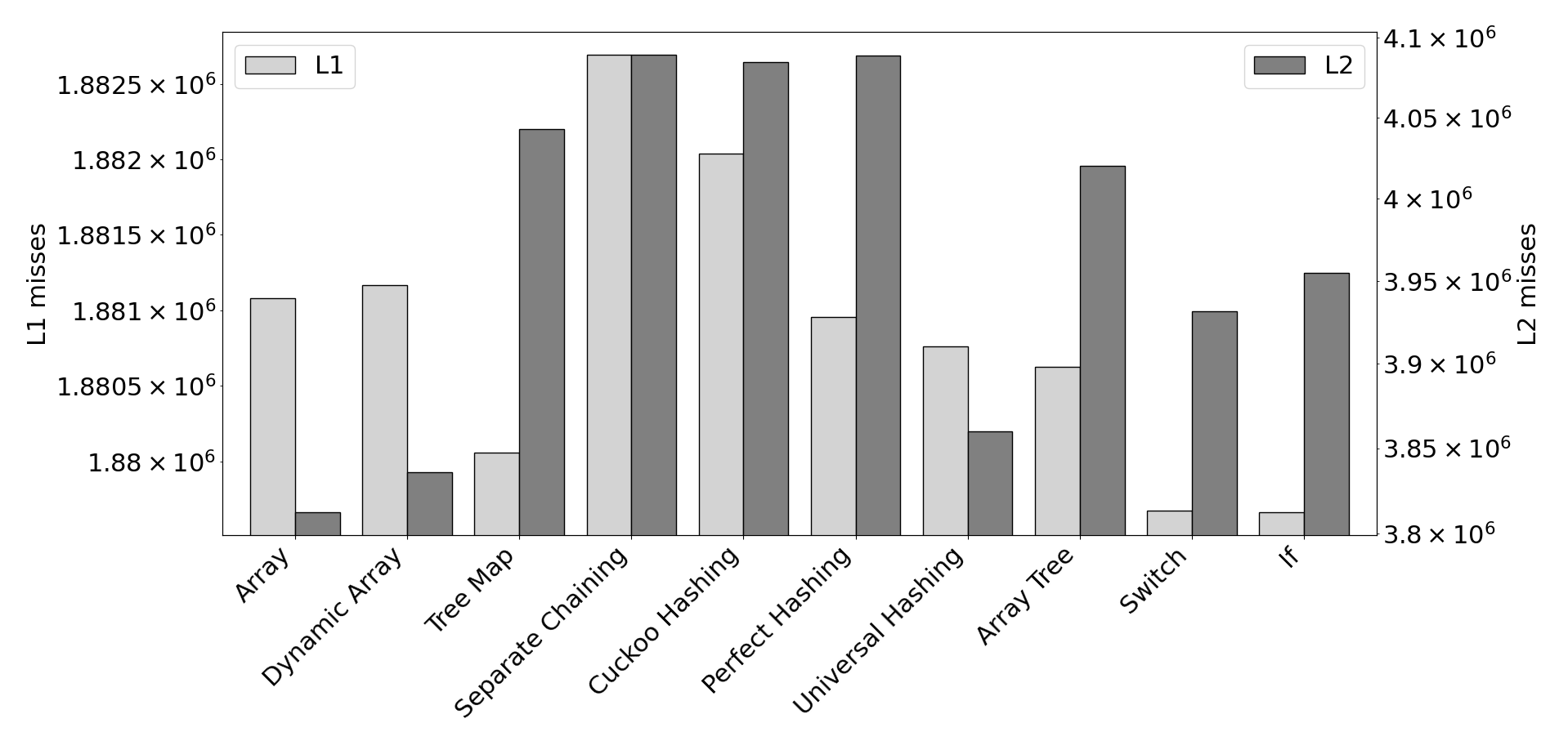}
  \caption{L1 and L2 cache misses using the trace based on the CIC-IDS17 dataset with an identifier mapping that corresponds to Zeek's default protocol support. Each bar corresponds to the average of 10 runs. Note that the differences in absolute values are small.}
  \label{fig:eval_cache}
\end{figure}

It is not surprising that the data structures based on more advanced hash functions come with a significant performance cost,
as can be seen in \Cref{fig:eval_zeek}.
Given the small set of involved identifiers,
the tree-based data structures perform better.
The best performance is achieved by the array approaches,
which may even outperform the hard-coded variants depending on their realization by the compiler.
Although it is a common approach to translate larger control-flow structures into jump tables that resemble arrays,
compilers may chose to realize different techniques like a binary search, with respect to the memory trade-off for sparse jump tables%
\footnote{While we observed significant differences between clang and GCC, both compilers allow for modification of their standard behavior.}.
This explains the relatively poor performance of the hard-coded variants for the randomized trace using the fragmented mapping as depicted in \Cref{fig:eval_fragmented}.
Likewise, the performance of the tree-based data structures foreseeably degrades for the fragmented mapping.
The universal hashing approach is able to keep up with the array approaches due to the simplicity of the employed hash function.

In terms of memory consumption, as expected, the tree-based and hash-based approaches scale well for increasing numbers of identifiers.
In contrast, the static array consumes a fixed amount of memory that remains largely unused.
For compact mappings, the memory consumption can be improved by using a dynamic array trimmed at front and end.
Yet, the memory utilization remains poor for widely scattered mappings.
However, assuming two byte identifiers and a 64 bit platform, the maximal size of an array would be $512$ KiB,
which is still negligible for common deployments.
With respect to the partially significant memory consumption,
we also consider the caching behavior.
In general, the handling of identifier mappings can be assumed to be cache-friendly due to their static nature.
\Cref{fig:eval_cache} shows the number of cache misses for the evaluated data structures%
\footnote{Given the peculiarities of the CPU architecture in terms of the relation between the hardware performance counters for different cache levels,
we focus on the comparison of the data structures per cache level.}.
Although the sparse arrays cause relatively high numbers of L1 data cache misses compared to other data structures,
the comparatively good performance for L2 cache misses suggest that their simple structure is ultimately beneficial for caching.
Even under the deteriorated conditions, the caching behavior does not compromise lookup performance of arrays.

Finally, startup time becomes relevant for the universal hashing approach.
While the other data structures can be built in just a few milliseconds,
finding a collision free hash function becomes disproportionately time consuming with the size of the identifier mappings increasing.
All in all, the advantages of arrays outweigh their fairly high memory consumption,
especially with regard to the predictability of the achieved performance.

\subsection{Monitor Performance}
\label{sec:Monitor_Performance}

Based on the previous results, we select the dynamic array for our implementation of the extended NSM architecture in Zeek.
The dispatching performance for the dynamic array is comparable to the hard-coded approaches.
Furthermore, the implementation is straight-forward, which supports maintainability.

To measure the impact on the monitor's overall performance,
we broke down the previously hard-coded analysis of packet-level protocols in Zeek into packet analyzers.
The migration to packet analyzers is discussed in more detail in \Cref{sec:traditional_stack}.
For both versions, the original hard-coded one and the modularized version,
we compared execution times and maximum memory usage for processing the CIC-IDS17 Monday PCAP, using Zeek's default configuration.
The results as depicted in \Cref{tbl:eval_zeek} show that the impact of the extended architecture is negligible with only minimal increases in processing time and memory usage.

\begin{table}[t!]
    \centering
    \begin{tabular}{l | c c} 
     & Runtime [s] & Memory [MB] \\
     \hline
     Original Zeek & $8.31$ & $108.46$ \\ 
     Extended Zeek & $8.39$ & $110.99$ \\
     \hline
     Difference [\%] & $~0.8$ & $~2.3$\\
    \end{tabular}
    \vspace{5pt}
    \caption{Comparison between original Zeek using hard-coded packet-level analysis and Zeek implementing the extended architecture,
    when processing the CIC-IDS17 Monday trace (1M packets).
    The numbers are the average of 10 runs.}
    \label{tbl:eval_zeek}
\end{table}

\section{Practical Applications}
\label{sec:practical_applications}

In this section we showcase the new possibilities offered by the modularization of packet-level analysis.
With respect to the requirements as defined in \Cref{sec:requirements_and_implementation},
we demonstrate the benefits of \emph{configurability} (R4) and illustrate the \emph{universality} (R2) of the extended NSM architecture on the example of Zeek's previously hard-coded analysis of packet-level protocols.
Furthermore, we extend the Spicy parser generator toolchain to facilitate the development of packet-level dissectors.
Then, we discuss two practical applications for monitoring industrial control systems,
presenting packet analyzers for GOOSE and ProfinetIO that exploit the \emph{extensibility} (R1) of the proposed architecture focusing on \emph{independency} of PA-Plugins (R3).
Finally, we introduce the logging of unknown protocols to increase visibility and prevent unnoticed monitor evasions in light of \emph{configurability} (R4).

\subsection{Traditional Protocol Stack Migration and Spicy Parser Generator}
\label{sec:traditional_stack}

As part of the implementation of the extended NSM architecture in Zeek,
we migrated Zeek's previously hard-coded processing of packet-level protocols.
In \Cref{sec:Monitor_Performance}, we showed that modularization introduces only a negligible performance overhead compared to the hard-coded version (R5).
In the following, we provide details on the realization of \emph{configurability} (R4) and demonstrate the \emph{universality} (R2) of the proposed architecture based on the migrated protocol stack.
Additionally, we discuss the extension of the Spicy parser generator to aid the development of new packet-level dissectors.

\begin{figure}[t!]
  \centering
  \includegraphics[width=0.9\columnwidth]{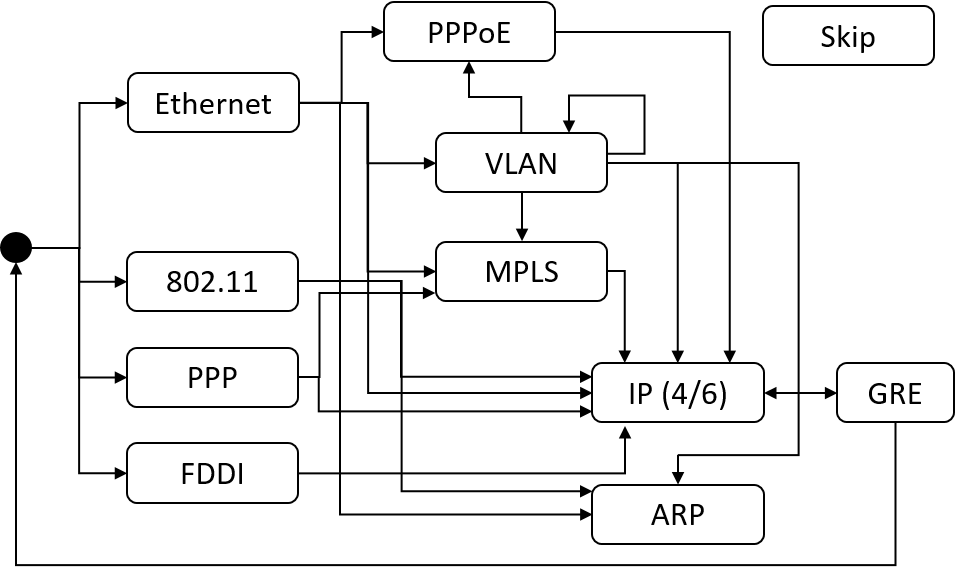}
  \caption{Protocol transitions of Zeek's traditional packet-level stack. The capture-related protocols are omitted for readability and handling of tunnels is simplified.}
  \label{fig:traditional_stack}
\end{figure}

\paragraph{Traditional Protocol Stack Migration}
Zeek's original implementation covers eleven packet-level network protocols:
Ethernet, FDDI, IEEE 802.11, PPP, PPPoE, VLAN, MPLS, ARP, IPv4, and IPv6 as well as GRE.
In addition, Zeek also supports a set of capture-related headers: LinuxSLL, NFLOG, BSD loopback encapsulation (NULL), and IEEE 802.11 RadioTap.
Furthermore, we add a special analyzer that allows for skipping a configured amount of bytes.
The skip analyzer enables NSM operators to quickly adapt to complex setups that exhibit fixed but unknown headers often introduced by proprietary protocols.
\Cref{fig:traditional_stack} shows the protocol transitions of the traditional packet-level protocol stack.
Employing the new framework for modular packet analysis,
the depicted transitions are entirely configurable by the NSM operator using policy scripts,
which act as Zeek's primary user interface.
\Cref{lst:ether_config} provides an excerpt from the PA-Config script for the Ethernet packet analyzer.
The script defines the transition table that maps protocol identifiers, in this case EtherTypes,
to the corresponding packet analyzers to be registered during initialization of Zeek.
Registration is done using the \lstinline{register_packet_analyzer} function that receives the parent analyzer (\lstinline{ANALYZER_ETHERNET}) to register a mapping for and an identifier together with the corresponding child analyzer%
\footnote{Note that Zeek 4.0 uses multiple calls to the registration function, omitting tables.}.
To customize the analysis process, operators only need to adapt the given transition tables.
Overall, the \emph{configurability} (R4) achieved by migrating Zeek's traditional protocol stack to the extended NSM architecture enables the adaption of the previously hard-coded analysis process as well as the reuse of the existing protocol analyzers in new contexts.

\begin{lstlisting}[float,floatplacement=H,
caption={PA-Config excerpt that shows the implementation of the transition table for the Ethernet packet analyzer.
The transition table maps EtherTypes to packet analyzers.
During initialization, the defined mappings are registered.},
emph={global, of, redef, &, event, for}, label=lst:ether_config]
global ether_types: table[count]
of PacketAnalyzer::Tag = {
    [0x8847] = PacketAnalyzer::ANALYZER_MPLS,
    [0x0800] = PacketAnalyzer::ANALYZER_IP,
    [0x86DD] = PacketAnalyzer::ANALYZER_IP,
    [0x0806] = PacketAnalyzer::ANALYZER_ARP,
    [0x8035] = PacketAnalyzer::ANALYZER_ARP,
    [0x8100] = PacketAnalyzer::ANALYZER_VLAN,
    [0x88A8] = PacketAnalyzer::ANALYZER_VLAN,
    [0x9100] = PacketAnalyzer::ANALYZER_VLAN,
    [0x8864] = PacketAnalyzer::ANALYZER_PPPOE
  } &redef;

event zeek_init() &priority=20
  {
  for ( id, child_analyzer in ether_types )
    PacketAnalyzer::register_packet_analyzer(
      PacketAnalyzer::ANALYZER_ETHERNET,
      id, child_analyzer);
  }
\end{lstlisting}

While we transfer the traditional stack without implementing additional parsing functionality,
the \emph{universality} (R2) of the proposed architecture allows us to account for previously unaddressed special cases.
For example, Zeek's original Ethernet implementation focused on Ethernet II frames only.
However, depending on the range the EtherType value falls into and the first two bytes of the payload,
Ethernet frames can be differentiated into Novell raw IEEE 802.3, IEEE 802.2 LLC, IEEE 802.2 SNAP, and Ethernet II frames.
By allowing the Ethernet analyzer to call configurable sub analyzers,
we employ the proposed architecture to offer possibilities for future extension.

In addition to the modularization of data-link and internet layer protocol dissectors,
the extended architecture also lays the foundation for the flexibilization of transport layer protocol dissectors that perform the transition from packet- to session-level.
With respect to the reference architecture, there are two options to realize this transition:
On the one hand, packet analyzers may implement protocol-specific session tracking by moving the session table logic into the packet analyzer itself.
On the other hand, since session tracking is a common task,
the monitor core may provide a generic session table component to be shared by multiple analyzers.
While out of scope for this work, the Zeek project has already begun work on the latter approach.

\paragraph{Spicy Parser Generator}
Though the refined NSM architecture allows for easy addition of new dissectors,
implementing the parsing logic still remains a cumbersome and error-prone process \cite{sassaman_halting_2011}.
As a network security monitor needs to deal with arbitrary network traffic from potentially malicious sources,
the robustness of packet analyzers is a major concern.
To further aid the development of packet-level dissectors,
we have added support for packet analyzers to the Spicy parser generator \cite{sommer_spicy:_2016}.
Spicy defines a language to integrate the specification of syntax and semantics for data formats ranging from network protocols to file formats and offers tooling to generate dissectors based on the format specifications.
Furthermore, the Spicy toolchain already integrates with Zeek.
\Cref{lst:spicy} shows the specification for a packet-level analyzer that parses IEEE 802.11Q VLAN tags.
The first 16 bits comprise the Tag Control Information (TCI), which is divided into the VLAN Identifier (VID), the Drop Eligible Indicator (DEI), and a Priority Code Point (PCP).
Then the actual EtherType is parsed and used to forward the remaining data to the next analyzer, once parsing the tag is done.
Spicy being able to translate high-level specifications into a pluggable dissector completely eliminates the need to write C++ code and thus vastly simplifies the development of packet analyzers.
Especially with respect to the binary nature of packet-level protocols, this approach also promises to yield more robust protocol parsers,
which is crucial for the overall security of the monitor.

\begin{lstlisting}[float,floatplacement=H,
caption={Functional Spicy example that specifies a Zeek packet analyzer for 802.1Q VLAN tags.},
emph={module, import, public, type, unit, on}, mathescape=false, label=lst:spicy]
module VLAN;
import zeek;
public type Packet = unit {
    tci: bitfield(16) {
        vid: 0..11;
        dei: 12;
        pcp: 13..15;
    };
    ether_type: uint16;

    on %done {
        zeek::forward_packet(self.ether_type);
    }
};
\end{lstlisting}

\subsection{Monitoring in Industrial Control Systems}
\label{sec:monitoring_ics}

In the context of Industrial Control Systems (ICS),
the integration of so-called Operational Technology (OT) and traditional Information Technology (IT) increases.
Connecting ICS to larger networks or even the internet introduces new attack vectors that represent substantial threats,
e.g., when exposing critical infrastructures \cite{hutchison_towards_2013, elbez_new_2018}.
State-of-the-art NSMs typically lack the ability to provide visibility into domain-specific ICS protocols.
In the following, we show how the proposed modular NSM architecture can be used to extend a network security monitor to make information extracted from ICS communication available for security operations by exploiting the \emph{independency} (R3) of PA-Plugins.
To this end, we extend Zeek with packet-level analyzers for GOOSE, used in the energy sector, and ProfinetIO, employed in production.
We briefly introduce each protocol, present attacks discussed in recent research,
and demonstrate the value of integrating modular packet-level dissectors by leveraging them to implement corresponding attack detection logic in Zeek's domain-specific scripting language.

\begin{lstlisting}[float, floatplacement=H,
caption={Configuration for the integration of GOOSE.},
emph={redef}, mathescape=false, label=lst:app_config]
redef PacketAnalyzer::ETHERNET::ether_types +=
{
  [0x88b8] = PacketAnalyzer::ANALYZER_GOOSE,
  [0x88b9] = PacketAnalyzer::ANALYZER_GOOSE
};
redef PacketAnalyzer::VLAN::protocols +=
{
  [0x88b8] = PacketAnalyzer::ANALYZER_GOOSE,
  [0x88b9] = PacketAnalyzer::ANALYZER_GOOSE
};
\end{lstlisting}

\paragraph{GOOSE Protocol}
\label{sec:goose}

The Generic Object-Oriented Substation Events protocol (GOOSE) is used for the communication between Intelligent Electronic Devices (IEDs) of electrical substations.
The protocol is standardized as part of IEC 61850.
To meet real-time requirements,
GOOSE builds directly on top of Ethernet and applies a multicast communication scheme.
Because GOOSE lacks authentication mechanisms,
the protocol is susceptible to machine-in-the-middle attacks.
Attackers with access to the substation network may inject false data to interfere with the operation of the power grid.

The need for visibility into GOOSE communication is well recognized.
\citeauthor{kabir-querrec_cyber_2017} already extended Zeek to parse GOOSE messages \cite{kabir-querrec_cyber_2017}.
Recently, another patch has been released by the ResiGate project of the Advanced Digital Sciences Center (ADSC)~\cite{chen_goose_2020}.
However, due to the far-reaching changes these extensions require to the monitor core,
the code was not officially integrated into Zeek.
The proposed extended architecture for modular packet-level analysis allows us to decouple the dissection of the GOOSE protocol and the monitor core,
by moving the existing code into a PA-Plugin%
\footnote{\url{https://github.com/kit-dsn/zeek-goose-analyzer}}.
Considering that GOOSE packets are identified by two EtherTypes (\lstinline{0x88b8} and \lstinline{0x88b9}) and can be found either directly in Ethernet frames or using VLANs, again,
the \emph{configurability} (R4) of the proposed architecture proves to be advantageous.
\Cref{lst:app_config} shows the corresponding configuration for GOOSE.
Note that the existing transition tables are extended using a Zeek language construct (\lstinline{redef}) that allows for the addition of new entries to existing tables.

Due to the limited practical feasibility of mechanisms to secure GOOSE communication,
a lot of recent research particularly addresses attack detection \cite{hoyos_exploiting_2012, kabir-querrec_cyber_2017, bohara_ed4gap_2020}.
For example, \citeauthor*{bohara_ed4gap_2020} discuss a so-called poisoning attack \cite{bohara_ed4gap_2020}:
During normal operation, GOOSE endpoints regularly send messages announcing their current state.
Each message also contains two counters, state number ($st$) and sequence number ($sq$).
For each repeated message $sq$ is increased.
In case of an event that changes the encoded state,
$st$ is increased to signal the state change.
After a state change, messages are sent in a higher frequency before exponentially slowing down to the original sending interval.
Based on the counter values, endpoints discard already processed messages to reduce load.
In a simple attack, valid messages can be "overwritten" by an attacker sending out messages with high counter values.
Monitoring the progression of the counter values, this type of attack can be easily detected.

\begin{lstlisting}[float, floatplacement=H, belowskip=-10pt,
caption={Exemplary Zeek script to detect $st' > st+1$},
emph={event, type, local, global, print}, mathescape=false, label=lst:goose]
global stNums: table[string] of count;

event goose_message(info: GOOSE::PacketInfo,
    pdu: GOOSE::PDU) {
  local ds: string = pdu$datSet;
  # Initialization
  if ( ds !in stNums ) {
    stNums[ds] = pdu$stNum;
    return;
    }
  # Check for increments > 1
  if ( pdu$stNum > stNums[ds] + 1 )
    print fmt("State number jump to %d",
      pdu$stNum);
  # Update counter
  stNums[ds] = pdu$stNum;
  }
\end{lstlisting}

Given the \emph{independency} (R3) of PA-plugins,
the modular GOOSE dissector is able to generate an event for each GOOSE message.
The event receives the message content including the counter values as an argument so that policies can be defined with respect to these values.
\Cref{lst:goose} shows a simplified example of an event handler that detects jumps of the state number ($st$) larger than one.
We track the counters per GOOSE data set in the \lstinline{stNums} table.
In case the data set has not been seen before, the counter values are initialized.
Otherwise, we check for increments larger one and print a message if we detect a jump.
Finally, the tracked counters are updated.
For brevity, we omit the handling of counter rollovers.
Furthermore, attackers may adapt to the currently valid counter numbers.
\citeauthor*{bohara_ed4gap_2020} present an advanced detection logic that detects inadequately overlapping resends,
which indicate the insertion of messages using valid counters by a third party \cite{bohara_ed4gap_2020}.
In this regard, the extended architecture for packet-level analysis promotes the transfer of current research into practice by establishing a clear separation of concerns.
On the one hand, dissecting the GOOSE protocol is decoupled from the monitor core and moved into an easy to deploy PA-Plugin to be maintained separately.
On the other hand, the demand-driven integration of the dissector in the overall monitoring system allows security experts to exploit the synergies of a unified interface,
as provided by the domain-specific Zeek scripting language,
to implement and share new detection approaches.

\paragraph{ProfinetIO Protocol}
\label{sec:profinet}

\begin{figure}[t!]
  \centering
  \includegraphics[width=0.65\columnwidth]{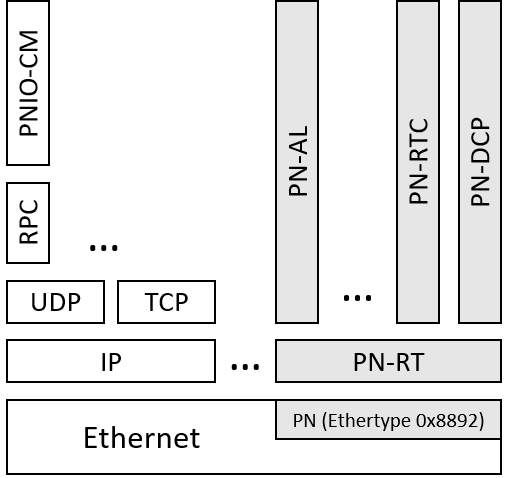}
  \caption{Profinet protocol suite (c.f. \cite{osadcii_design_2017}). IP-based protocols are supported while the Ethernet-based stack is unsupported by traditional NSMs.}
  \label{fig:profinet_stack}
\end{figure}

In a typical production scenario a process is implemented using field devices such as sensors and actuators that are controlled by Programmable Logic Controllers (PLCs).
ProfinetIO handles the communication between PLCs and field devices as well as other components such as programming stations or human-machine interfaces.
With more than 32 million deployed nodes \cite{weber_profinet_2020},
ProfinetIO is one of the most widely used Ethernet-based fieldbus protocols and was standardized as part of IEC 61158.
The Profinet protocol suite (c.f. \Cref{fig:profinet_stack}) can be divided into several protocols and sub-protocols.
In particular, the following three protocols need to be considered:
The Discovery and Configuration Protocol (DCP), the Context Management protocol (CM), and the Real Time Cyclic protocol (RTC).
During the start up phase of a Profinet system, PLCs use the Ethernet-based DCP to discover field devices by name and provide an IP configuration for these devices.
Subsequently, the UDP-based CM protocol can be used to configure process-related communication relations between devices and controllers.
Once the setup is completed, the actual process control communication is handled by the RTC protocol using one of three priority classes according to the previously configured relations.
While existing NSMs can be readily extended to monitor the UDP-based CM communication%
\footnote{For example, Amazon's \citetitle{amazoncom_inc_zeek_2021} \cite{amazoncom_inc_zeek_2021}},
there is no visibility into the larger part of the stack that is Ethernet-based.

\citeauthor{pfrang_detecting_2018} investigate replay-attacks against the Ethernet-based ProfinetIO \cite{pfrang_detecting_2018}.
To successfully replay traffic, they disrupt the existing communication relationship between a field device and its controlling PLC by conducting a reconfiguration attack.
In a reconfiguration attack DCP is used to rename a victim device, which terminates existing connections.
Afterwards, the attacker can establish new communication relations to execute a subsequent replay attack.
By monitoring the Ethernet-based DCP, these attacks can be easily detected as well.

Leveraging the extended NSM architecture for modular packet-level analysis,
we implement an analyzer for ProfinetIO focusing on DCP.
DCP uses the common PN-RT header that is transferred over Ethernet and identified by EtherType \lstinline{0x8892}.
Again, we exploit the \emph{independency} (R3) of PA-plugins to enable the definition of DCP-specific policies by generating events that reflect the request-response-oriented communication flow.
For our implementation we employ the previously introduced Spicy parser generator,
which allows for the specification of Zeek events to be triggered based on the progress of parsing.
To detect successful renaming attacks, we track the configured values of the observed devices to distinguish between benign rewrites and malicious changes of device names.
This is done by correlating renaming requests, responses that confirm renaming, and subsequent searches for the old names.
We provide both, the ProfinetIO packet analyzer and the detection script, in a separate Zeek plugin%
\footnote{\url{https://github.com/kit-dsn/zeek-profinet-analyzer}}.

With respect to the complexity of the Profinet protocol stack that is characterized by the interactions and dependencies of the different protocols,
being able to gather and correlate information from the various subsystems offers new possibilities for comprehensive, system-scope analyses.
Yet simple, the example attack detectors for GOOSE and ProfinetIO clearly demonstrate the value of modular packet analyzers.
The separation of policy and mechanism (c.f. \cite{paxson_bro:_1999}) using a domain-specific language for monitoring facilitates the access of security experts to new application areas and promotes the exchange and transfer of existing knowledge.
In addition to detecting known attack patterns,
NSM operators might choose to apply environment-specific policies that flag deviations from expected behavior.
Furthermore, machine learning approaches could be applied to implement anomaly detection.
In this context, we would like to emphasize again that we see our work as complementary to the use of machine learning methods,
since the quality of these methods significantly depends on the selection and preprocessing of input data \cite{anderson_machine_2017}.

\subsection{Unknown Protocols}
\label{sec:unknown_protocols}

Visibility is paramount in network security monitoring, but, as explained in \Cref{sec:background},
it is neither feasible nor desirable to support monitoring of all protocols.
Because protocol parsing itself is complex, unnecessary dissectors increase the monitor's attack surface in terms of implementation errors.
Furthermore, attackers might be able to degrade the monitor's performance by confronting an NSM with network traffic that is specially crafted to trigger expensive processing using otherwise irrelevant protocols.
Once protocol identification becomes ambiguous, the simultaneous operation of multiple protocol dissectors might even be exploited to evade the monitor \cite{grashofer_attacks_2020}.
However, making use of \emph{configurability} (R4) by just excluding protocols would immediately introduce a new attack vector,
if attacks using these protocols go completely unnoticed.

To prevent excluded protocols from being exploited to bypass the monitor,
we log the use of unknown protocols.
In the \emph{unknown\_protocols.log}, the time, the analyzer that encountered an unknown protocol, the corresponding protocol identifier, and a configurable amount of bytes from the unknown PDU are logged.
If a protocol unknown to the NSM is regularly used in the monitored network,
logging each unknown packet would quickly overwhelm the monitor.
Thus, we also introduced a throttling mechanism.
For each protocol analyzer, the packets per unknown protocol identifier are counted.
Once the threshold is reached, the reporting of packets using this unknown protocol identifier is rate-limited to a given sampling rate.
The rate-limiting expires after a specified duration.
Threshold, sampling rate, and rate-limiting duration are configurable as well.

We expect the logging of unknown protocols to significantly improve the deployment of an NSM in new environments.
By evaluating the unknown protocols log, NSM operators can quickly determine if additional protocols need to be supported or spot misconfigurations.
In a parallel effort during our work, ESnet\footnote{\url{https://www.es.net/}}
already proved the value of unknown protocol detection in practical operations.

\section{Conclusion}
\label{sec:conclusion}

With the increasing proliferation of information technology in new areas such as energy, mobility, and production,
NSMs need to cope with the corresponding diversification of the TCP/IP-oriented protocol stack.
While existing approaches for modularization of protocol dissection at application layer provide a high degree of flexibility,
they do not extend to packet-level analysis,
leaving established NSMs incapable of dynamically integrating non-IP protocol dissectors.
Given that processing at these low levels is highly performance critical,
the question arises whether it is practically feasible to replace the hard-coded path with a dynamic plugin scheme.

In this paper, we present a refined software architecture for NSMs that enables the flexible integration of lower-layer protocol dissectors.
We investigate the challenges in the context of packet-level dispatching and derive detailed requirements for the implementation of the proposed, extended architecture.
Instead of the prevalent notion of well-ordered protocol stacks,
in practice the interrelations between protocols form more complex protocol transition graphs.
With dispatching performance being of utmost importance,
we find that hash-based data structures introduce a significant overhead,
whereas array-based approaches can keep up with the hard-coded processing at packet-level.
Based on this result, we implement the extended NSM architecture in the open-source NSM Zeek,
migrate Zeek's previously hard-coded stack of packet-level protocols to the new interface, and verify that the performance impact is indeed negligible.
We demonstrate the benefits of the proposed architecture by implementing basic attack detection techniques as proposed in recent research on the security of industrial control systems in the fields of energy and production,
employing the new interface to realize modular dissectors for GOOSE and \mbox{ProfinetIO}, two popular industrial communication protocols.
To further assist the practical application of modular packet-level protocol dissectors,
we integrate with the Spicy parser generator, which allows for creating new dissectors in a declarative fashion,
and present a feature to increase visibility when a monitor is confronted with unknown protocols.

With our work, we hope to support the scientific community as well as practitioners.
In light of the rapid evolution and increasing adoption of information technology into ever new environments,
we seek to extend visibility for security operations,
facilitate the adoption of existing technologies and, in particular,
promote the transfer of application-domain-driven research into practice.
Thus, all artifacts are made publicly available including the implementation of the extended NSM architecture for modular packet-level analysis being part of Zeek's 4.0 release.

\section*{Acknowledgments}
Hannes Hartenstein and Jan Grashöfer acknowledge the funding of the Helmholtz Association (HGF) through the Competence Center for Applied Security Technology (KASTEL).

\printbibliography

\end{document}